\UseRawInputEncoding
\documentclass[pdflatex,sn-mathphys-num]{sn-jnl}

\usepackage{graphicx}%
\usepackage{multirow}%
\usepackage{amsmath,amssymb,amsfonts}%
\usepackage{amsthm}%
\usepackage{mathrsfs}%
\usepackage[title]{appendix}%
\usepackage{xcolor}%
\usepackage{textcomp}%
\usepackage{manyfoot}%
\usepackage{booktabs}%
\usepackage{algorithm}%
\usepackage{algorithmicx}%
\usepackage{algpseudocode}%
\usepackage{listings}%
\usepackage{array}
\usepackage[caption=false,font=normalsize,labelfont=sf,textfont=sf]{subfig}
\usepackage{stfloats}
\usepackage{url}
\usepackage{verbatim}
\usepackage{hyperref}
\usepackage{soul}
\usepackage[normalem]{ulem}

\theoremstyle{thmstyleone}%
\theoremstyle{thmstyletwo}%

\theoremstyle{thmstylethree}%

\begin{document}

\title[Article Title]{Frequency drift corrected ultra-stable laser through phase-coherent fiber producing a quantum channel}


\author[1]{\fnm{Stanley} \sur{Johnson}}\email{stanley@iucaa.in}

\author[2]{\fnm{Sandeep} \sur{Mishra}}\email{sandeep.mishra@jiit.ac.in}

\author[2]{\fnm{Anirban} \sur{Pathak}}\email{anirbanpathak@yahoo.co.in}

\author*[1]{\fnm{Subhadeep} \sur{De}}\email{subhadeep@iucaa.in}

\affil[1]{\orgname{Inter-University Centre for Astronomy and Astrophysics (IUCAA)}, \orgaddress{\street{Post bag 4, Ganeshkhind}, \city{Pune}, \postcode{411007}, \state{Maharashtra}, \country{India}}}

\affil[2]{\orgdiv{Department of Physics and Materials Science and Engineering}, \orgname{Jaypee Institute of Information Technology}, \orgaddress{\street{A 10, Sector 62}, \city{Noida}, \postcode{201309}, \state{Uttar Pradesh}, \country{India}}}



\abstract{Phase coherent fibers (PCF) are essential to distribute nearly monochromatic photons, ultra-stable in their frequency and phases, which have demanding requirements for state-of-the-art networked experiments, quantum as well as very high-speed communications. We report the development of a novel system that produces PCF links, also actively corrects the unavoidable slow frequency drift of the source laser. The PCF follows white phase noise limited $\sigma_o \times \tau^{-1}$ stability behavior having $\sigma_o$ values $1.9(2) \times 10^{-16}$ and $2.6(1) \times 10^{-16}$ for a 3.3 km field-deployed and 71 km spool fibers, respectively, with up to 47.5 dB suppression of the phase noise compared to a normal fiber. Additionally, the system is featured to correct the source laser's 33.8 mHz/s frequency drift to as low as $\simeq 0.05$ mHz/s. Therefore, this all-in-one solution producing a quantum link can potentially enhance the effectiveness of the twin field quantum key distribution (TF-QKD) by nearly a 73-fold reduction of the QBER that arises from using unstabilized fiber links\textcolor{black}{, as well as relaxes the laser frequency drift correction constraints by severalfold}.}




\maketitle

\section{Introduction}\label{sec1}
Phase coherent optical fiber-based transfer of nearly monochromatic, and frequency cum phase stabilized optical signals are inevitable for wide ranges of geographically distributed yet networked experiments, for example, time and frequency metrology \cite{calonico2017italian, tampellini2019coherent, boulder2021frequency}, fundamental physics tests through inter-comparison of accurate sensors \cite{safronova2018two, oelker2019demonstration, brady2023entanglement}, geodetic measurements \cite{mehlstaubler2018atomic}, astronomy \cite{krehlik2017fibre}, quantum teleportation \cite{shen2023hertz, krutyanskiy2023entanglement}, quantum communication \cite{lucamarini2018overcoming, pittaluga2021600, pittaluga2025long, liu2023experimental, zhou2023twin, clivati2022coherent} and many more. For this, phase stabilized optical fibers are an unavoidable requisite in lab-scale \cite{ma1994delivering, 10.1063/5.0138599, jiang2010nd}, metro-scale \cite{foreman2007coherent, liu2023high} and long-haul \cite{predehl2012920, PhysRevA.92.021801, droste2013optical} links to transfer reference photons having $\leq10^{-17}$ fractional frequency stability \cite{huntemann2012high, oelker2019demonstration, bloom2014optical} through them. Stability of the optical phase, when it is used as the quantum basis, plays a crucial role in quantum key distribution (QKD) protocols \cite{berrevoets2022deployed, wang2022sub, eriksson2019wavelength}. In particular to the twin field quantum key distribution (TF-QKD) \cite{lucamarini2018overcoming, pittaluga2021600, liu2023experimental, zhou2023twin, clivati2022coherent, wang2020optimized} simultaneous communication of the `signal' laser at wavelength $\lambda_s$  through two independent long-haul fiber-based quantum channels, however, maintaining a \textcolor{black}{high stability of their} differential phase is \textcolor{black}{extremely} important to minimize the quantum bit error rate (QBER) \cite{pittaluga2021600}. \textcolor{black}{In practice, this phase difference is dynamic which originates from two different phenomena:  frequency drift of the lasers and instability of the optical fibers.} Therefore, in addition to the use of low-loss fibers that enhance the quantum bit rates and the communication length, the incorporation of fiber phase stabilization to suppress the phase drift by nearly four orders of magnitude results in significant improvement of QBER. Due to technical limitations, to overcome the frequency indistinguishable Rayleigh back-scattered photons from the signal photons, another `reference' wavelength $\lambda_r$ is also required \cite{pittaluga2021600}. A differential drift $<$1 Hz between the two signal lasers is a necessity for efficient performance of the TF-QKD. Therefore, the incorporation of a laser frequency drift correction system together with fiber phase stabilization certainly improves the TF-QKD.

In this work, we present a combined solution to generate end-to-end optically stabilized phase coherent fiber (PCF) inclusive of laser frequency drift correction of an ultra-stable laser and its transmission through the PCF. Reference optical resonator-stabilized ultra-stable lasers experience a slow, unidirectional frequency drift ($\leq$100 mHz/s) due to ageing of the resonator's material resulting in a change of its characteristics \cite{sterr2009ultrastable}. \textcolor{black}{The frequency drift between two arbitrarily drifting lasers has been demonstrated to be suppressed to 0.5 mHz/s \cite{liu2024creation} and 2 mHz/s \cite{zhou2024independent}.} A few different ways to measure the drift, such as, precision atomic spectroscopy \cite{degenhardt2005calcium}, measuring the differential drift between the mutually orthogonal dual axes in a reference cavity \cite{hill2021dual} have been studied and the drift suppression is reported to be as low as 0.5 mHz/s \cite{hill2021dual}. Optical fibers have also been used in the frequency stabilization of lasers \cite{kefelian2009ultralow, jiang2010agile, dong2015subhertz, kong2015long}. Here, we describe a novel optical self-referencing method for estimation followed by correction of the drift, and an absolute optical referencing cum correction technique that results in better suppression of drift than earlier reported results. The experimental setup, the low-noise servo and the signal processing are described in Sec. 2, the obtained results together with its applications in TF-QKD are presented in Sec. 3, followed by a conclusion in Sec. 4.

\section{Methods}\label{sec11}

\begin{figure}[!t]
\centering
\includegraphics[width=3.5in]{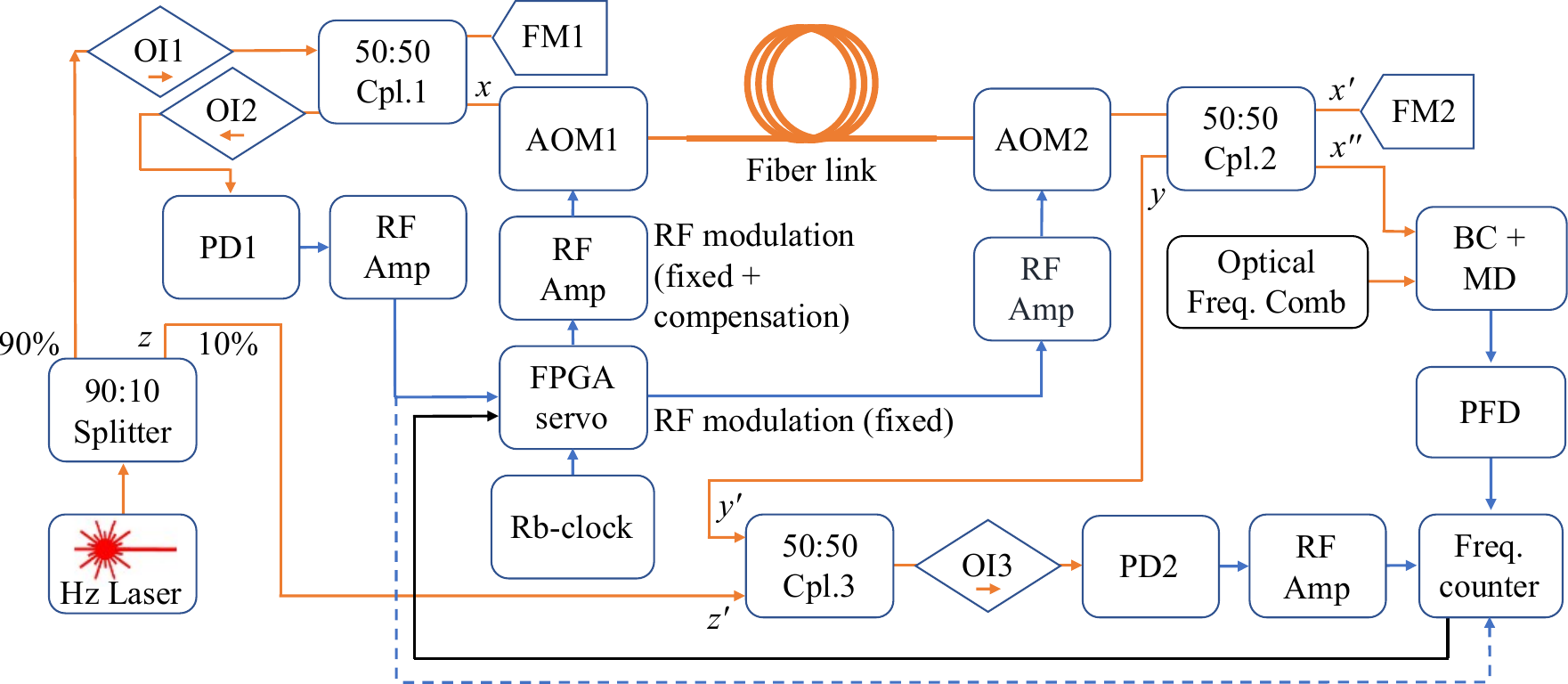}
\caption{The experiment setup consisting of Hz-laser: ultra-stable laser, OI1, 2 and 3: optical isolators, 50:50 Cpl.1, 2 and 3: 50:50 optical couplers, AOM1 and 2: acousto-optic modulators, FM1 and 2: Faraday mirrors, PD1 and 2: photodiodes, BC + MD: beam-combiner and monochromatic detector, PFD: phase-frequency detector, RF Amp: RF amplifier. The optical and electrical signals are shown in orange and blue, respectively. The black signal depicts the flow of Hz-laser drift information in the absolute optical frequency referencing case.}
\label{fig_1}
\end{figure}

\begin{figure}[!t]
\centering
\includegraphics[width=3.15in]{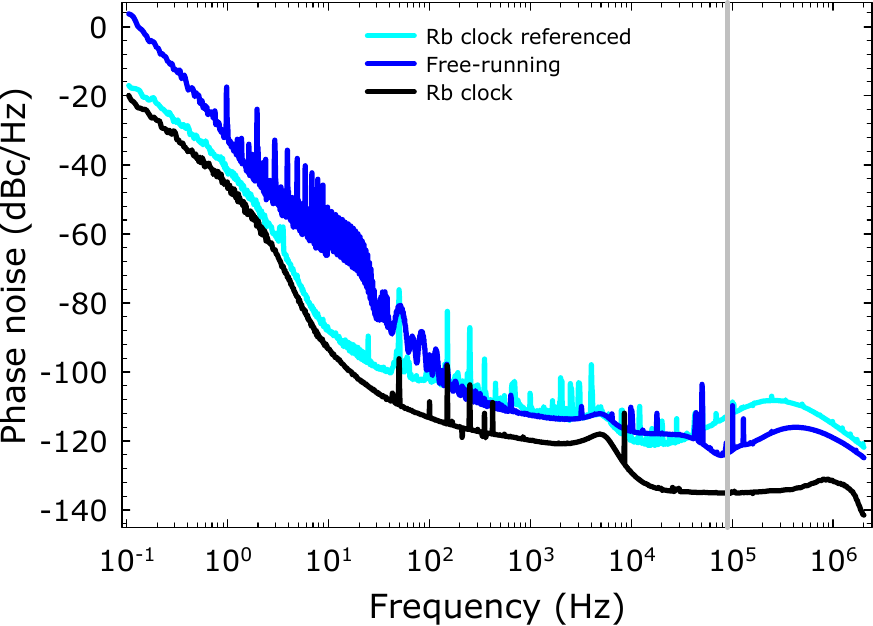}
\caption{Phase noise for the fixed frequency output of the FPGA at 20 MHz, for Rb clock referenced (cyan) and free-running (blue) cases, and the same are compared to the Rb clock (black). The gray vertical line indicates the PID servo open loop unity gain bandwidth, i.e. 90 kHz.}
\label{fig_3}
\end{figure}

Figure \ref{fig_1} shows the setup to produce the PCF link, which is an end-to-end optical fiber guided system. An ultra-stable laser with about 1 Hz linewidth (Hz-laser) operating at 1550 nm wavelength is used as the reference source. Its output is split in a 90:10 ratio; the 10$\%$ light propagates through a 1 m fiber (\textit{z$\rightarrow$z'}) and is used as the nearly unperturbed optical frequency reference for out-of-loop beat signal generation at the optical coupler Cpl. 3 that is detected by photodiode PD2. The 90$\%$ light is further equally split into two: one half of it is transported through an acousto-optic modulator (AOM1), the link fiber, and AOM2 (\textit{x$\rightarrow$x'}), whereas the other half is retro-reflected by the Faraday mirror (FM1) to use as the reference light. Both the AOMs operate at 20 MHz center frequency (\textit{f\textsubscript{1}} and \textit{f\textsubscript{2}}) with $\pm$1 MHz bandwidth. The AOM1 incorporates an active frequency correction corresponding to the optical fiber phase noise \textit{$\phi_{F}^{n}(t)$} and Hz-laser's frequency drift, whereas AOM2 distinguishes the Rayleigh backscattered photons produced in the fiber from the retro-reflected light by FM2, in frequency space. Optical isolators OI1-3 suppress reflected photons returned to the system, therefore reducing the possibility of forming any unwanted optical resonator. Efficient optical isolation plays a crucial role since the PCF uses bi-directional optical transmission, and is, therefore, very sensitive to reflection-induced noise. The round-trip \textit{x}$\rightleftharpoons$\textit{x'} light carries twice the fiber link noise, given as, 

\begin{equation}
f_{F}^{n} = (2\pi)^{-1} \, d\phi_{F}^{n}/dt,
\end{equation}
\label{eq_1}

\noindent which together with the intrinsic drift of the laser frequency, is determined by beating the retro-reflected light with the reference Hz-laser at PD1 (in-loop beat signal) and sent to an in-house developed field programmable gate array (FPGA) based servo, that generates a proportionate frequency correction and applies it to the AOM1 for active stabilization to create the PCF. The servo hardware consists of a lock-in amplifier, a frequency measurement unit, two proportional-integral-derivative controllers PID 1, 2, and a variable frequency synthesizer, all implemented digitally using a low-noise FPGA board \cite{johnson2025atomic}, which is synchronized to the 10 MHz output of a GPS referenced Rubidium clock (Rb clock) that also serves as a common stable external frequency reference for all parts of the PCF setup. 
This makes the hardware operation low-noise, for example, Fig. \ref{fig_3} depicts an 85$\%$ suppression of the phase noise of our FPGA generated outputs at 20 MHz (used for \textit{f\textsubscript{1}} and \textit{f\textsubscript{2}}) with the benefit of low long-term frequency drift when Rb clock referenced, compared to a free-running condition. 

\begin{figure*}
\centering
\includegraphics[width=6.50in]{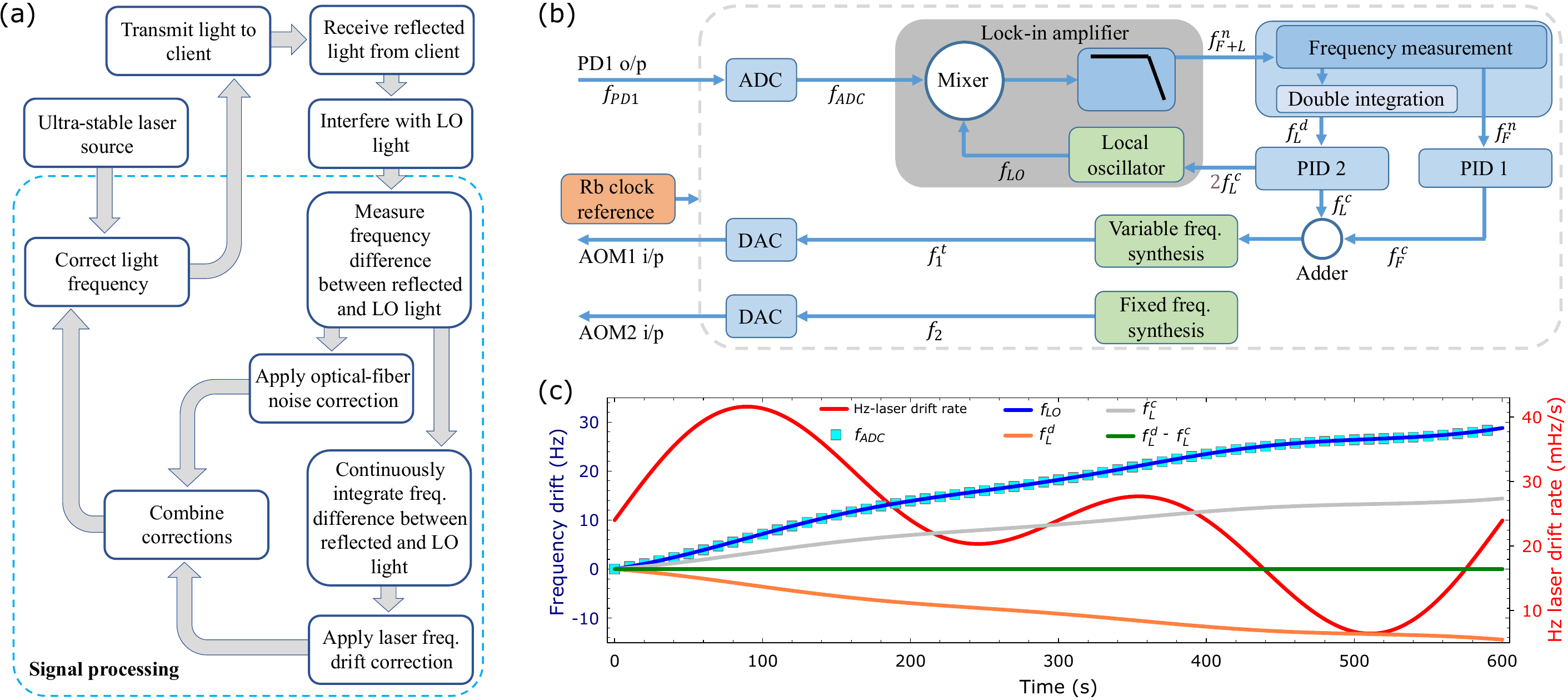}
\caption{The optical self-referencing technique: (a) Flowchart for the signal processing, (b) Schematic of the signal processing for the system, ADC: Analog-to-digital converter, DAC: Digital-to-analog converter, PID 1 and 2: Proportional-integral-derivative controllers. (c) Simulated example of the working of the signal processing system for source laser drift compensation, with the uncompensated Hz-laser drift rate (red) referenced to the Y axis on the right. All other curves are referenced to the Y axis on the left and depict frequency drift for: uncompensated Hz-laser \textit{$f_{L}^{d}$} (orange), in-loop beat signal \textcolor{black}{after digitization \textit{$f_{ADC}$}} (cyan squares), local oscillator \textit{$f_{LO}$} (blue), compensation signal \textit{$f_{L}^{c}$} (gray) and compensated Hz-laser \textit{$f_{L}^{d}$} - \textit{$f_{L}^{c}$} (dark green).}
\label{fig_2}
\end{figure*}

The Hz-laser has a slow frequency drift \textcolor{black}{$f_L^d$ at a rate typically} $< 50$ mHz/s due to the ageing of its reference Fabry–Pérot cavity that is used for frequency stabilization and linewidth narrowing of the laser \cite{zhadnov2018thermal}. For measuring this drift at the remote end of the fiber link, an out-of-loop beat signal between the Hz-laser and the 1550 nm output of an RF stabilized optical difference frequency comb (DFC) is generated using a beam combiner, grating filter, and fast photodetector, BC + MD, (see Fig. \ref{fig_1}). A phase-frequency detector (PFD) further filters and amplifies the beat signal, which is then recorded on an Rb clock referenced precision frequency counter \cite{KpK_website}.  

We have implemented a novel optical self-referencing technique to correct the drift of the laser frequency. Figure \ref{fig_2} (a, b) depicts the signal processing flowchart and the corresponding electronic arrangement, respectively, for optical self-referencing. For this, the PD1 acquires the fractional drift of the laser frequency $\Delta \textit{$f_{L}^{d}$}$, together with the fiber link noise (Fig. \ref{fig_1}), with the in-loop beat signal given as,
 \begin{equation}
f_{PD1} = 2(f_{1} + f_{2} + f_{L}^{c} + f_{F}^{n}) + \Delta f_{L}^{d}.
\label{eq_2}
 \end{equation}
Here, \textit{$f_{L}^{c}$} is the servo generated compensation frequency for the laser's drift, and the factor of 2 arises since the retro-reflected light passes twice through \textit{x}$\rightleftharpoons$\textit{x'} for the in-loop beat detection. \textit{f\textsubscript{PD1}} is around 80 MHz, primarily due to the dominance of $2(f_{1} + f_{2})$. The ADC digitizing \textit{f\textsubscript{PD1}} has a sampling rate \textit{f\textsubscript{S}} = 125 MSa/s, therefore operating in the second Nyquist zone \cite{TI_Nyquist_zones} and the beat signal carrier frequency after digitization is given by, 
\begin{align}
\begin{split}
f_{ADC} = f_S - f_{PD1},
\end{split}
\label{eq_3}
\end{align}

\noindent which is in the vicinity of \textit{f\textsubscript{S} - 2(f\textsubscript{1} + f\textsubscript{2})} = 45 MHz. The local oscillator therefore produces the reference RF for lock-in detection given by, 
\begin{align}
\begin{split}
f_{LO} = [f_S - 2(f&_{1} + f_{2})] - 2f_{L}^{c}.  
\end{split}
\label{eq_4}
\end{align}
The low-pass filter is set to pass the signal at around \textcolor{black}{$(f_{ADC} - f_{LO})$} $\leq 113$ kHz, and the lock-in amplifier detects the frequency noise present in the in-loop beat signal (Eq. \ref{eq_2}), given as,
 \begin{equation}
f_{F+L}^{n} = \textcolor{black}{f_{ADC} - f_{LO}} = \textcolor{black}{-}[2f_{F}^{n} +\Delta f_{L}^{d}].  
\label{eq_5}
\end{equation}
%
$\Delta \textit{$f_{L}^{d}$}$ has one of its components from the frequency drift of the Hz-laser \textit{$f_{L}^{d}$} and the other is due to slow variation of temperature changing the \textcolor{black}{\textit{x}$\rightleftharpoons$\textit{x'}} optical path length. Hence, double integration is employed for the separation of \textit{$f_{F}^{n}$} and the suppression of temperature variations in the estimation of \textit{$f_{L}^{d}$} from $\Delta \textit{$f_{L}^{d}$}$. Thereafter, \textit{$f_{F}^{n}$} and \textit{$f_{L}^{d}$} are servoed through PID 1 and 2, respectively (Fig. \ref{fig_2} (b)). For optical self-referencing, we employed self-heterodyning, where the retro-reflected light passes through the PCF link that acts as an optical delay line. This retro-reflected light is used to interfere with the Hz-laser. Therefore, the in-loop beat measures the frequency drift over the delay time, which is possible since the link fiber is phase stabilized.   

\begin{figure}[b]
\centering
\includegraphics[width=3.5in]{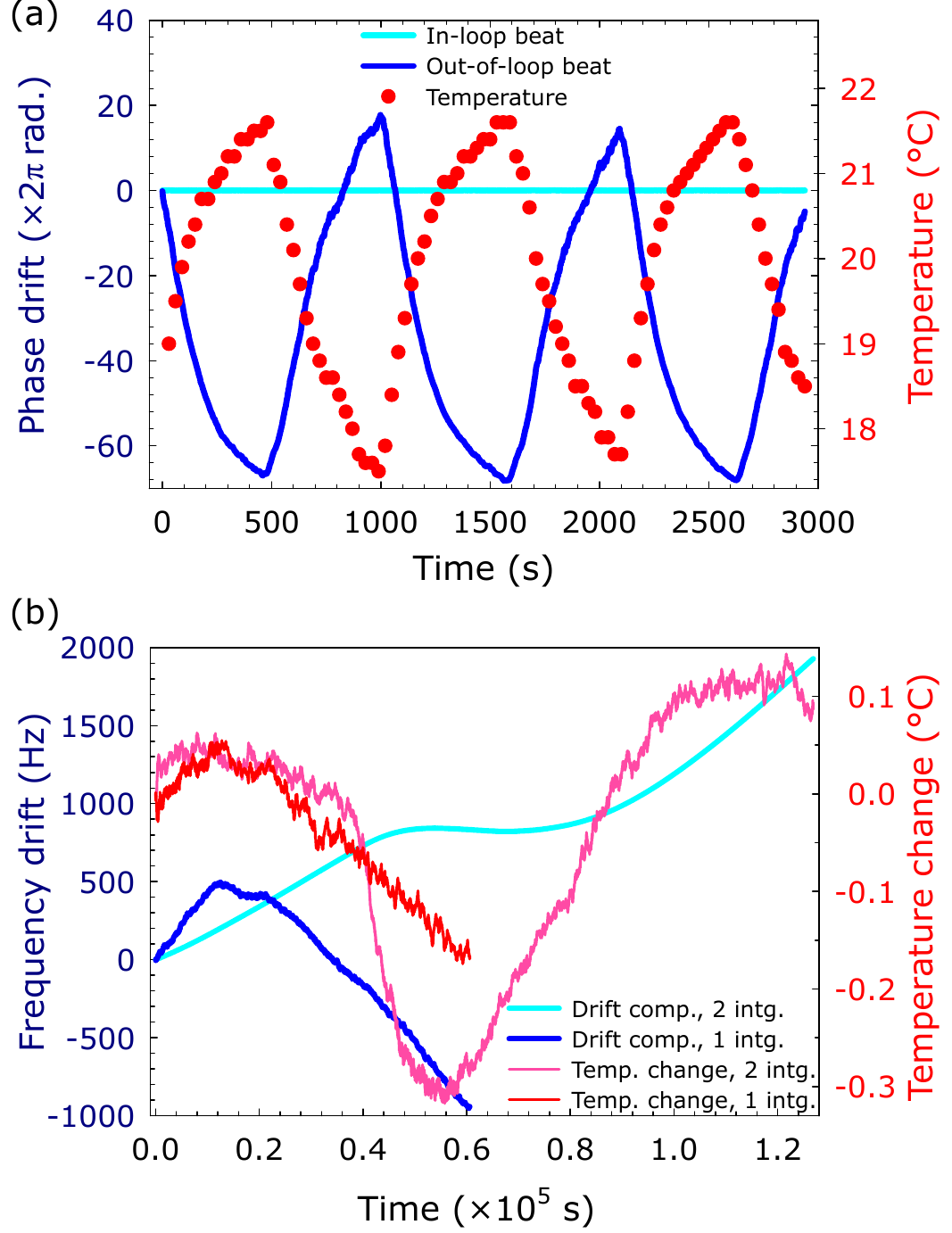}
\caption{(a) Phase drift for a 1 m fiber link with fiber noise cancellation active, for the in-loop beat signal (cyan) and out-of-loop beat signal (blue), the temperature in the vicinity of the experiment is depicted in red. (b) AOM1 frequency drift compensation signal \textit{$f_{L}^{c}$} (blue, left Y axis) drifting with temperature (red, right Y axis) for a 5 m fiber link, when only one of the two integrators before PID 2 are active, and \textit{$f_{L}^{c}$} (cyan, left Y axis) not drifting with temperature (pink, right Y axis) for the 5 m fiber link, when two integrators before PID 2 are active.}
\label{fig_4}
\end{figure}

The PID 2 also actuates the local oscillator frequency \textit{f\textsubscript{LO}} (see Eq. \ref{eq_4} and Fig. \ref{fig_2} (b)) and its continuous adjustment provides a dynamically stable reference for subsequent frequency drift measurement, therefore allowing for \textit{$f_{F}^{n}$} and \textit{$f_{L}^{d}$} to be simultaneously corrected. This unique feature incorporated in our setup makes it more efficient by not only generating a phase coherent fiber link but also simultaneously suppressing the laser's frequency drift. The outputs of both PID 1 and 2 that represent the frequency compensation to be applied to AOM1 are then combined and given to the variable frequency synthesis block, which generates the AOM1 driving frequency, 
\begin{align}
f_{1}^{t} = f_{1} - f_{F}^{c} - f_{L}^{c}. 
\label{eq_6}
\end{align}

\noindent Here, \textit{$f_{F}^{c}$} is the compensation frequency for fiber link noise. Figure \ref{fig_2} (c) shows simulated waveforms depicting various signals involved in the signal processing. For this, we considered an example time variation of the Hz-laser's frequency, resulting in \textit{$f_{L}^{d}$}. Thereafter the local oscillator adjusts \textit{f\textsubscript{LO}}, following the PID's action, to closely match the drift in the in-loop beat note \textcolor{black}{after digitization \textit{f\textsubscript{ADC}}} ($\approx$ 2\textit{$f_{L}^{c}$}), and the Hz-laser light is transmitted through the PCF with compensation \textit{$f_{L}^{d}$} - \textit{$f_{L}^{c}$}. A combined action of all the interconnected elements in the signal processing (Fig. \ref{fig_2} (b)) facilitates the real-time measurement and compensation for (i) the fiber link noise and (ii) the drift of the Hz-laser frequency.

Figure \ref{fig_4} (a) shows the temperature variations affecting the \textit{x$\rightarrow$x'}, \textit{y$\rightarrow$y'} and \textit{z$\rightarrow$z'} fibers, causing changes in phase of the in-loop and out-of-loop beat signals, when PCF is active for the link fiber \textit{x$\rightarrow$x'} (Fig. \ref{fig_1}). The servo actively compensates phase deviations for the in-loop signal, whereas the out-of-loop beat signal's phase mimics the temperature variations. Therefore, the system is useful for sensitive monitoring of temperature. For the objective of this report, the measurement of frequency drift necessitates the compensation of this temperature impact. A single integrator before PID 2 gives inefficient performance, therefore an additional integrator is implemented and their effectiveness is depicted in Fig. \ref{fig_4} (b). A single integrator causes \textit{$f_{L}^{c}$} to follow temperature changes. By employing the second integrator, the correlation between \textit{$f_{L}^{c}$} and temperature is broken, thus enabling better compensation.

\section{Results and discussion}

\begin{figure}[t]
\centering
\includegraphics[width=5.5in]{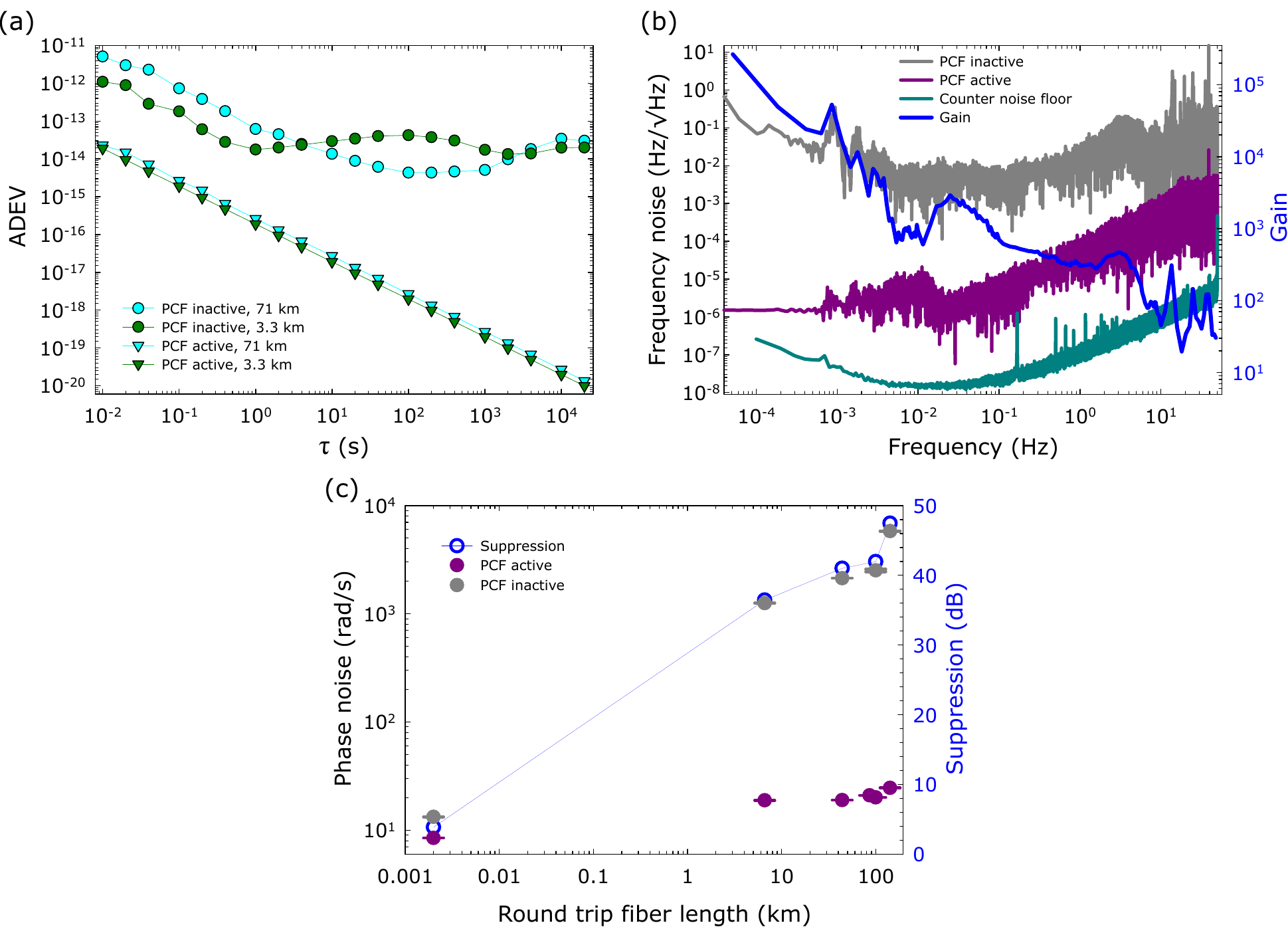}
\caption{(a) Allan deviation (ADEV) of the fractional frequency instability for the in-loop beat signal for a fiber spool of length 71 km (\textcolor{black}{cyan}) and for a 3.3 km deployed fiber link (dark green). Circles and \textcolor{black}{triangles} depict the ADEV when the PCF is inactive and active, respectively. (b) The Y-axis on the left depicts the frequency noise spectrum of the in-loop beat signal for a 50 km fiber spool when the PCF is active (purple) and inactive (gray); frequency counter noise floor (dark cyan). The Y-axis on the right depicts the corresponding gain of the system (blue). (c) The Y-axis on the left depicts the phase noise of the in-loop beat signal for various lengths of fiber when the PCF is active (purple) and inactive (gray), and the Y-axis on the right depicts the corresponding phase noise suppression (blue). The laser frequency drift compensation was inactive for (a), (b) and (c).}
\label{fig_5}
\end{figure}

The Allan deviation (ADEV) of the fiber links: geographically distributed, deployed underground, and using spools \textcolor{black}{within an environment-controlled lab that are protected from the direct path of air currents coming from air-conditioning vents}, is shown in Fig. \ref{fig_5} (a) for both PCF active and inactive conditions, when the laser frequency drift compensation was inactive. For these, the feedback signal from PD1 is used to measure the stability of the fiber links. The system is tested for up to 71 km of fiber spools and also with a 3.3 km fiber link loop (lab$\rightarrow$out of lab$\rightarrow$lab) deployed on campus. The ADEV for the PCF links follow $\sigma$($\tau$) = $\sigma_{o} \times \tau^{-1}$ behaviour for an averaging time $\tau$, indicating the system is white phase noise dominated \cite{Benkler_2015}. We obtained $\sigma_{o}$ to be $1.9(2)\times10^{-16}$ and $2.6(1)\times10^{-16}$ for the underground deployed and spool fibers, respectively, which are lower than the reported values of $4\times10^{-16}$ for a 110 km fiber \cite{musha2008coherent} and $7\times10^{-15}$ for a 235 km fiber \cite{cizek2022coherent}. We measure an integrated phase noise of 6.9 m rad for our 3.3 km deployed PCF link, which is better than the 18.2 m rad for a 7 km deployed fiber estimated from the reported results in Ref. \cite{foreman2007coherent}.

For example, Fig. \ref{fig_5} (b) compares the frequency noise spectrum for 50 km unstabilized and PCF links. Therefore, it is clearly evident that there are nearly five orders of magnitude noise suppression at 1 mHz and a factor of 172 overall for the PCF link, which allows the transport of ultra-stable, nearly monochromatic laser light in a phase coherent manner without losing its characteristics. The frequency noise in an unstabilized fiber rises below $10^{-2}$ Hz, which worsens the ADEV for $\tau>10^{2}$ s as depicted in Fig. \ref{fig_5} (a). For better clarity, we also show the measured gain provided by our servo system, which effectively suppresses dominant acoustic, seismic, and thermal fluctuations that result in fiber instability. The phase noise for various lengths of unstabilized and PCF links, along with corresponding noise suppressions, is shown in Fig. \ref{fig_5} (c). A maximum of 47.5 dB suppression is achieved, which nearly saturates, even though the phase noise in the unstabilized fibers increases with the increasing fiber lengths as expected. Therefore, Fig. \ref{fig_5} (c) clearly demonstrates the \textcolor{black}{phase noise suppression performance of the developed system over different link lengths}, which is finally limited by \textcolor{black}{the} available laser power.  

\begin{figure}[!t]
\centering
\includegraphics[width=3.15in]{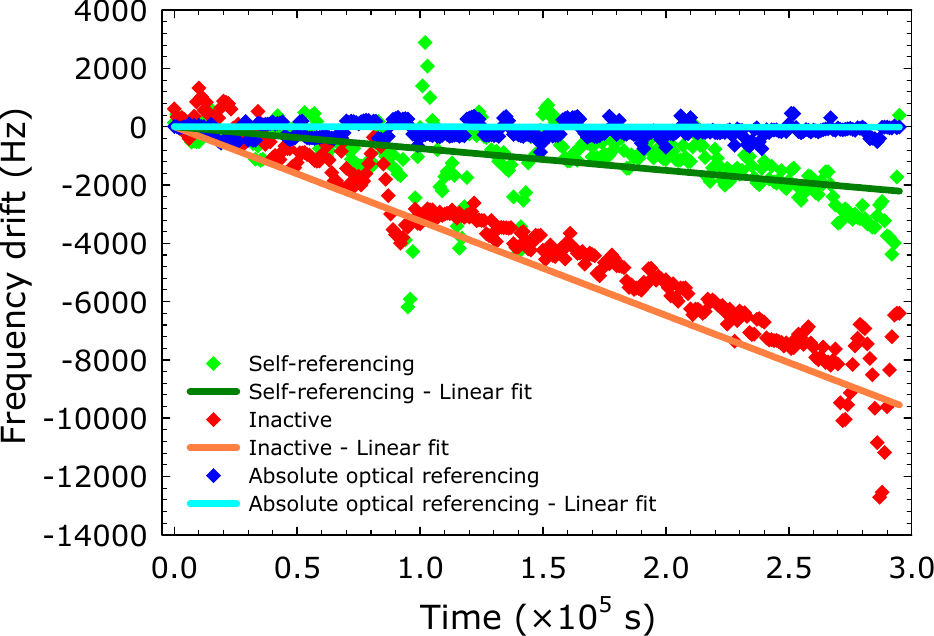}
\caption{Frequency drift of the beat between the light at the remote end and the light from the Rb clock referenced optical DFC, with Hz-laser drift compensation active (self-referencing: green symbols, absolute optical frequency referencing: blue symbols) and inactive (red symbols). The lines depict the linear fit for drift compensation active (self-referencing: dark green, absolute optical frequency referencing: cyan) and inactive (orange) cases.}
\label{fig_6}
\end{figure}

The effectiveness of optical self-referencing is measured by the out-of-loop beat signal between the Hz-laser and the DFC, which acts as an optical frequency reference. The DFC has a fractional frequency instability of $8.40(1) \times 10^{-12}$ at 1 s with each output line having a linewidth of 6.38 kHz when phase locked to an Rb clock. To detect the Hz-laser's drift but not the short-term variations in the beat frequency, averaging over a day is necessary because the DFC has lower stability in the hours timescale. In self-referencing, the PID 2 parameters are adjusted to minimize the drift. Since the laser drift is independent of the fiber length, we tested the system with a 5 m fiber and recorded the beat signal for $>3$ days. Figure \ref{fig_6} depicts the recorded data with and without activating the self-referencing drift compensation when the PCF was active for both cases. An average long-term drift rate is estimated from the slopes of linear fits to the data, which amounts to 6.2(9) mHz/s and 33.8(1) mHz/s upon employment of self-referencing and when free running, respectively. This shows an 82$\%$ compensation in the Hz-laser's frequency drift, therefore boosting its usefulness for various precision applications.  

Optical self-referencing is cost-effective since it can be implemented without an absolute optical frequency reference, such as a DFC, which we use only for monitoring and is not part of the feedback loop, however, the residual effect of temperature on the fiber limits the drift compensation. We also investigated an absolute optical referencing technique that uses the beat note between the PCF link transported Hz-laser and the DFC to estimate the drift in \textcolor{black}{an averaging interval of 24 hours} and compensates accordingly. We use an Extensible Messaging and Presence Protocol (XMPP)-based communication link \cite{XMPP_website} to transmit measured drift from the frequency counter to the FPGA servo, which accordingly gives an active correction to AOM1. Figure \ref{fig_1} depicts the experimental arrangement, where a direct connection from the frequency counter to the FPGA servo (black line) was established instead of the PD1$\rightarrow$RF Amp$\rightarrow$FPGA servo connection (that was used for optical self-referencing). The PCF i.e. PID 1 is still active in this technique, however, the double integration and PID 2 are not needed as the drift compensation works in a closed loop based on drift measurement by a counter, and hence is not estimated by the FPGA servo. This measured drift information received by the FPGA servo is used to generate the desired $f_{LO}$ and $f_{1}^{t}$ for Hz-laser drift compensation. The obtained result is shown in Fig. \ref{fig_6} corresponding to a drift of 0.05(12) mHz/s, which is a 99.9$\%$ reduction in the Hz-laser's frequency drift. Therefore, absolute optical frequency referencing is undoubtedly superior to optical self-referencing, however, its implementation is expensive as it requires a reference optical system which in our case is the DFC. Depending on an application's stringent needs, one may choose either of these reported drift compensation techniques.

\begin{table}[t]
\centering
\caption{Measured specifications of the system}
\label{tab:my_label}
    
    \begin{tabular}{>{\raggedright}p{0.02\textwidth}>{\raggedright}p{0.25\textwidth}|>{\raggedright\arraybackslash}p{0.15\textwidth}}
    \toprule
         \textbf{ } & \textbf{Parameters} & \textbf{Specifications}\\ 
         \hline 
         (i) & ADEV of PCF at $\tau$ = 2000 s & $1(3)\times10\textsuperscript{-19}$ (3.3 km `deployed' fiber), $1.35(5)\times10\textsuperscript{-19}$ (71 km fiber `spool'),
         $4.97(7)\times10\textsuperscript{-20}$ (1 m fiber `residual')\\
         (ii) & $\sigma_{o}$ & $1.9(2)\times10\textsuperscript{-16}$ (deployed), $2.6(1)\times10\textsuperscript{-16}$ (spool),  
         $8.6(1)\times10\textsuperscript{-17}$ (residual)\\   
         (iii) & Phase noise of PCF [rad/s]& 9.43(1) (deployed), 12.29(1) (spool)\\ 
         (iv) & Servo unity gain bandwidth (open loop)& 90 kHz\\
         (v) & Noise suppression by PCF (max.)& 47.5 dB\\
         (vi) &Suppressed laser frequency drift (3 days) [Hz/s] by:&  \\
          & $\cdot$ Optical self-referencing& $-6.2(9)\times10\textsuperscript{-3}$ \\
          & $\cdot$ Absolute optical referencing& $-5(12)\times10\textsuperscript{-5}$ \\
          \botrule
    \end{tabular}
\end{table}

\begin{figure}[!t]
\centering
\includegraphics[width=3.15in]{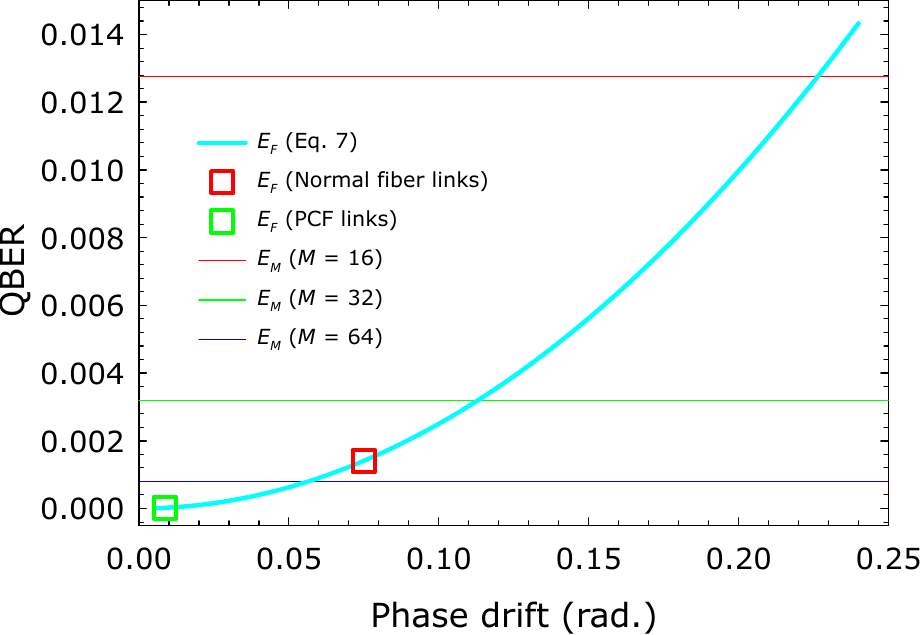}
\caption{Variations of QBER $E_F$ with the differential phase error $\delta \Phi$ between Alice-to-Charlie and Bob-to-Charlie links (cyan), for normal fiber links (red square), and PCF links (green square). The horizontal lines depict intrinsic QBER $E_M$ for $M =$ 16 (red), 32 (green), and 64 (blue).}
\label{fig_7}
\end{figure}

The PCF link has the potential to bring further sophistication in various state-of-the-art classical and quantum technologies, e.g., TF-QKD \cite{lucamarini2018overcoming, pittaluga2021600, pittaluga2025long}, quantum entanglement between geographically distant locations \cite{stolk2025extendable}, and time-frequency-phase dissemination with unprecedented accuracy \cite{lopez2013simultaneous, foreman2007coherent}. Here, we discuss its application in reference to the TF-QKD\textcolor{black}{, where the fiber links can be upgraded to PCF links implementing the described scheme in the same fiber}. Lucamarini \textit{et al.} \cite{lucamarini2018overcoming} developed TF-QKD to scale up the distance over which the optical fiber based quantum communication can be implemented without using any quantum repeaters. In TF-QKD, the signals are encoded within the global phase of $\lambda_s$ that is sliced to $2\pi/M$ parts, using a dim optical pulse for quantumness of the photons at both the remote ends, namely `Alice' (A) and `Bob' (B), which are then transmitted to a common central node `Charlie' (C) via two independent long haul fibers of nearly equal lengths. At Charlie's port, the pulses at $\lambda_s$ arriving from Alice and Bob at a given time $t$ are being interfered to detect the differential phase $\Phi = [\phi_{ss}^{AC}(t) - \phi_{ss}^{BC}(t)]$ between the signals \cite{suppl_info}, \textcolor{black}{where an error $\delta\Phi$ in estimating $\Phi$} originates due to their unequal optical path lengths A$\rightarrow$C and B$\rightarrow$C, resulting from non-identical fiber lengths and un-correlated noise coupled to the fibers leading to their different levels of instability. This signal pulse-to-pulse varying $\Phi$ perturbs encoded phase slice of $\lambda_s$ resulting in QBER by an amount of \cite{weihs2001photon}
 \begin{equation}
E_{F} = sin^{2}(\delta\Phi/2). 
\label{eq_7}
\end{equation}
\noindent This adds to the intrinsic value of QBER originating from $2\pi/M$ phase slices \cite{lucamarini2018overcoming}
 \begin{equation}
E_{M} = \frac{1}{2}\bigg[1 - \frac{sin(2\pi/M)}{2\pi/M} \bigg]. 
\label{eq_8}
\end{equation}

Considering the scheme of TF-QKD described by Liu \textit{et al.} \cite{liu2023experimental}, however, upon implementation of the PCF fiber links, as described in this article, instead of uncompensated fibers used in Ref. \cite{liu2023experimental}, we have estimated that $\delta\Phi$\textcolor{black}{, for an integration time of 40 $\mu$s as reported in Ref. \cite{liu2023experimental},} can be reduced from $4.3^{\circ}$ to $0.5^{\circ}$ for the same length of fibers \cite{suppl_info}. This results in the reduction of $E_F$ to $0.019\times10^{-3}$ from $1.4\times10^{-3}$ estimated using $\delta \Phi = 4.3^{\circ}$ reported in Ref. \cite{liu2023experimental}, as shown in Fig. \ref{fig_7}. \textcolor{black}{This reduction in the value of $\delta\Phi$ and hence reduction in the QBER can be useful for the enhancement of distance in TF-QKD.} Further, the reported optimal value of M = 16 \cite{lucamarini2018overcoming}, which corresponds to $E_M$ = $12.7\times10^{-3}$, can be fine sliced to 32 or even 64 if other technical challenges are possible to overcome, since their corresponding $E_M$ values $3.2\times10^{-3}$ and $0.8\times10^{-3}$, respectively, are reduced yet significantly dominate compared to $E_F$ = $0.019\times10^{-3}$ upon using the PCF links. \textcolor{black}{It is important to mention that we have not explicitly discussed the role of integration time relating QBER and the phase error while using TF-QKD as one of the use cases of PCF. To know more details on the role of integration time in TF-QKD, the readers can refer to Refs. \cite{clivati2022coherent,maruyama2025space}}. The described system, which, on the one hand, restricts the laser to sub-mHz/s frequency drift and also executes strong phase compensation of the fiber links, will have potential for the betterment of the TF-QKD.

\section{Conclusion}
We demonstrate the generation of a phase stabilized coherent optical fiber link using the in-house developed optical and electronic hardware that is capable of suppressing the phase noise by up to 47.5 dB compared to an unstabilized normal fiber. The developed system can simultaneously compensate the slow frequency drift of an ultra-stable source laser to 6.2 mHz/s using optical self-referencing and as low as 0.05 mHz/s using absolute optical frequency referencing techniques. The optical self-referencing does not require any additional reference or extra hardware compared to the PCF implementing system, making frequency drift reduction cost-effective and straightforward to implement. On the other hand, using an external optical frequency reference, such as an optical frequency comb that we used here, further improves the drift reduction performance of the system, as expected. The PCF incorporated with the optical frequency drift reduction, all implemented in one system that we have reported in this article, can help to deliver phase- and frequency-stable ultranarrow linewidth laser light over lab- and metro-scale fiber links, which has increasing demands for the very high-speed optical communication as well as for using them as quantum channels. For the twin field quantum key distribution (TF-QKD), in particular, we estimate a possibility of a 73-fold reduction in the quantum channel assisted quantum bit error rate (QBER) with the utilization of PCF links, thus improving the performance and scaling up of the communication distance of the TF-QKD. 



\backmatter

\bmhead{Supplementary information}
Supplementary Information: Frequency drift corrected ultra-stable laser through phase-coherent fiber producing a quantum channel: We estimate the phase noise reduction in TF-QKD upon using phase-coherent fiber.

\bmhead{Acknowledgements}

We thank the Dept. of Science and Technology, India, for their funding support through the QuEST programme.

\section*{Declarations}

\subsection*{Data availability}
All data generated or analysed during this study are included in this published article.

\subsection*{Author contribution}
S.J. conceived the idea for self-referencing drift compensation that is done in conjunction with the PCF, built the system, performed the experiments, and analyzed the results. S.M. and A.P. contributed to understanding the application of the PCF for TF-QKD. S.D. conceived the experimental idea, analyzed the results, performed theoretical calculations, and supervised overall. All authors drafted and reviewed the manuscript.

\bibliography{sn-article.bib}

\end{document}